\documentclass[twoside]{elsart} 
 \usepackage{epsfig,subfigure} 
 \newcommand{\SMALLCAP}	[1]	{\caption[]{\begin{scriptsize} #1 \end{scriptsize}}} 
 \newcommand{\ie}               {i.e.}  
   
 \newcommand{\bea}		{\begin{eqnarray}} 	
 \newcommand{\eea}		{\end{eqnarray}} 
 \newcommand{\beann}		{\begin{eqnarray*}} 	
 \newcommand{\eeann}		{\end{eqnarray*}}

 \newcommand{\lrb}		{\left(} 
 \newcommand{\rrb}		{\right)}

 \newcommand{\ve}		{\varepsilon} 
 \newcommand{\btpfuture}	{{\sc btp}--future} 
 \newcommand{\bundfuture}	{{\sc bund}--future} 
 \newcommand{\tbond}		{{\sc t--bond}} 
 \newcommand{\btp}		{{\sc btp}} 
 \newcommand{\bund}		{{\sc bund}} 
 \newcommand{\sbtp} 		{\mbox{{\scriptsize {\sc btp}}}} 
 \newcommand{\sbund} 		{\mbox{{\scriptsize {\sc bund}}}} 
 \newcommand{\ssbtp} 		{\mbox{{\tiny {\sc btp}}}} 
 \newcommand{\ssbund} 		{\mbox{{\tiny {\sc bund}}}} 
 \newcommand{\ds}		{\displaystyle} 
 \date{29 December 1998}
 
\begin{document} 
 \begin{frontmatter} 
 \title				{Correlations in the Bond--Future Market} 
 \author[INFM,Dresden]		{Gianaurelio Cuniberti\thanksref{email}}, 
 \author[INFM,Genova]		{Marco Raberto}, and 
 \author[INFM,Alessandria]	{Enrico Scalas} 
 \thanks[email]			{On leave of absence from Dipartimento di Fisica, Universit\`a di Genova, Italy; 
 				e--mail: {\tt cunibert@mpipks-dresden.mpg.de}; url:     
				{\tt www.infm.it/econophysics}} 
 \address[INFM]			{Istituto Nazionale per la Fisica della Materia, Unit\`a di Genova} 
 \address[Dresden]		{Max--Planck--Institut f\"ur Physik komplexer Systeme, \\ 
 				N\"othnitzer Stra{\ss}e 38, D-01187 Dresden, Germany} 
 \address[Genova]		{Dipartimento di Fisica, Universit\`a di Genova, via Dodecaneso 33, I-16142 Genova, Italy}   
 \address[Alessandria]		{Dipartimento di Scienze e Tecnologie Avanzate, 
 				Universit\`a del Piemonte Orientale, via Cavour 84, I-15100 Alessandria, Italy} 
 \begin{abstract} 
 We analyze the time series of overnight returns for the \bund \ and \btp \ futures exchanged at {\sc liffe} (London).   The overnight returns of both assets are mapped onto a one--dimensional symbolic--dynamics random walk: The ``bond walk".   During the considered period (October~1991---January~1994) the \bundfuture \ market opened earlier than the \btpfuture \ one.   The crosscorrelations between the two bond walks, as well as estimates of the conditional probability, show that they are not independent; however each walk can be modeled by means of a trinomial probability distribution.  Monte Carlo simulations confirm that it is necessary to take into account the bivariate dependence in order to properly reproduce the statistical properties of the real--world data.   Various investment strategies have been devised to exploit the ``prior" information obtained by the aforementioned analysis.  
 \end{abstract} 
 \end{frontmatter} 
 \section{Introduction} 
 Among social and economical disciplines, the analysis of financial markets is particularly suitable for a rigorous mathematical formulation.   
 More important, technological advances in computer science applied to financial trading make great amounts of data available.
 It is therefore possible, with great reliability, to match real--world information with theories, conjectures, and hypotheses, thus falsifying them in the spirit of the scientific method.
 Indeed, financial time series are the outcome of a many--agent interaction: The realm of statistical physics.
 It is not a surprise that nowadays an increasing number of physicists is working on problems of statistical finance \cite{AAP88}, \cite{Dumbar98}.
 One of the problems of practical interest is to investigate the existence of correlations between different asset time series \cite{Mantegna97b}.
 In principle, this information could be used in order to make profits, in practice, this possibility is almost always cancelled by transaction costs.
 \\ Here, we present a simple method to determine whether two financial time series are correlated.
 In particular, we have analyzed the time series of \bund \ and \btp \ futures exchanged at the London International Financial Futures and options Exchange ({\sc liffe}), during the period October~1991---January~1994, when the \bundfuture \  market opened earlier than the \btpfuture \ one.
 The overnight returns of both assets are mapped onto a one--dimensional symbolic--dynamics random walk: The ``bond walk" \cite{Scalas98}.
 \\ The paper is structured as follows: In section~\ref{sec:anres} we introduce the analysis tools and present the results.
 Section~\ref{sec:gamb} is devoted to the exploration of possible investment strategies using the information contained in correlations.
 Finally, in section~\ref{sec:concl} we draw our conclusions.
 \section{Analysis and Results} \label{sec:anres} 
\begin{figure} \begin{center} \subfigure[]{\epsfig{file=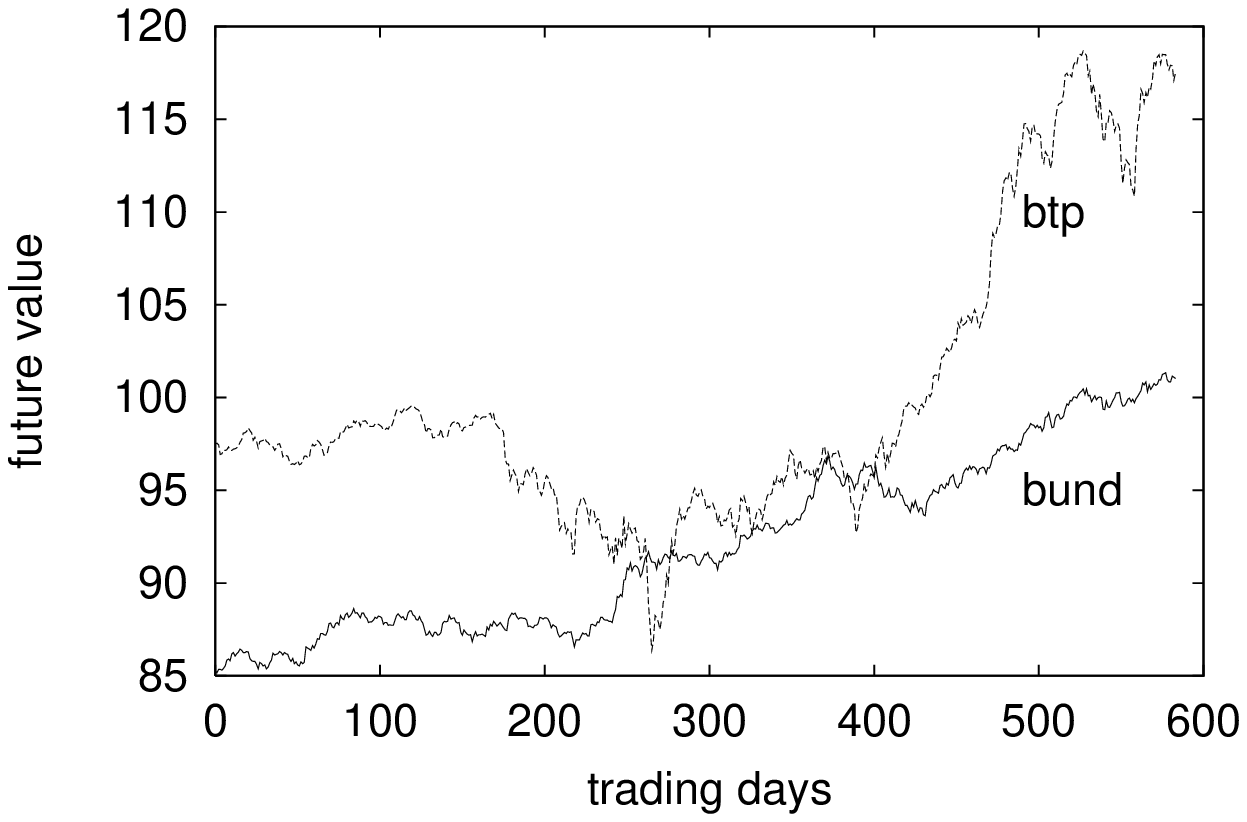, width=15pc}} \subfigure[]{\epsfig{file=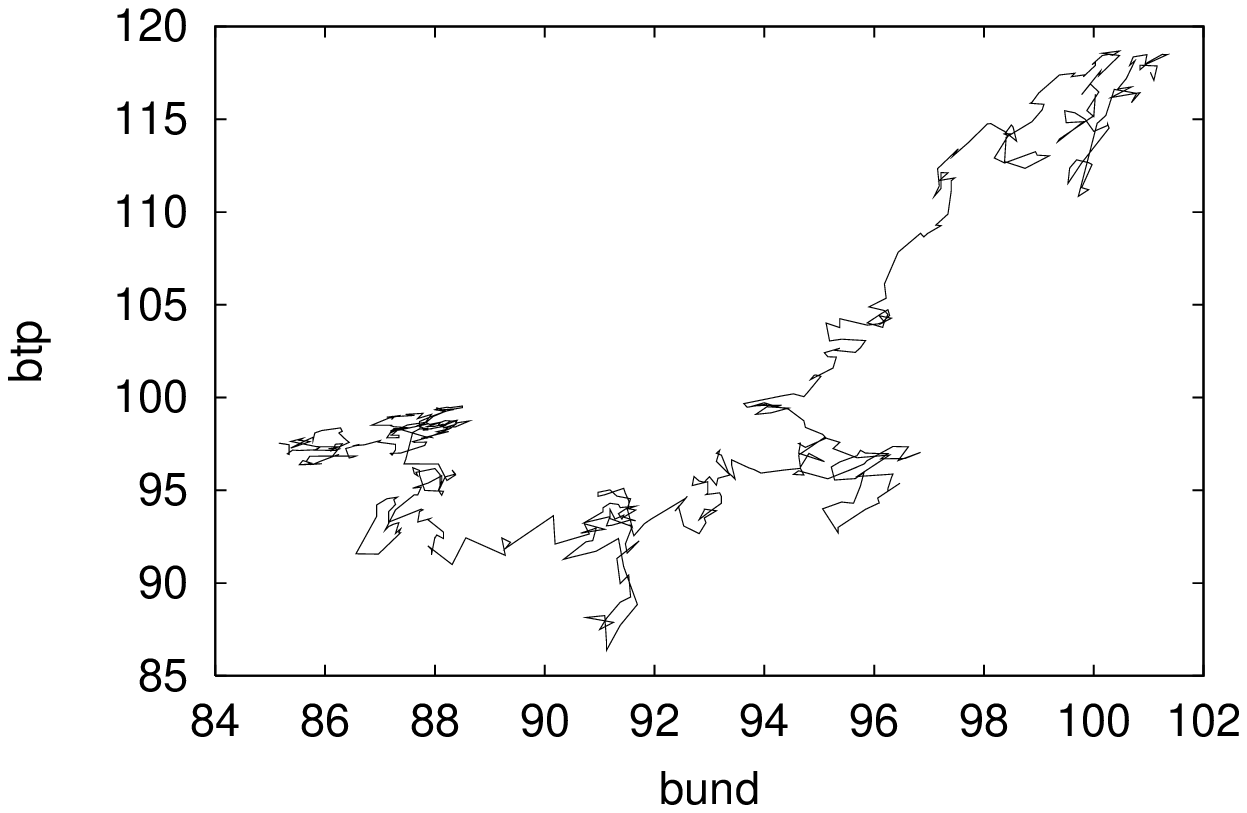, width=15pc}} \end{center} \SMALLCAP{(a) Closing prices of \bund \ (solid line) and \btp \ (dashed line) futures; (b) \btp--
 vs \bundfuture \ random walk} \label{fig:price} \end{figure}

\begin{figure} \begin{center} \subfigure[]{\epsfig{file=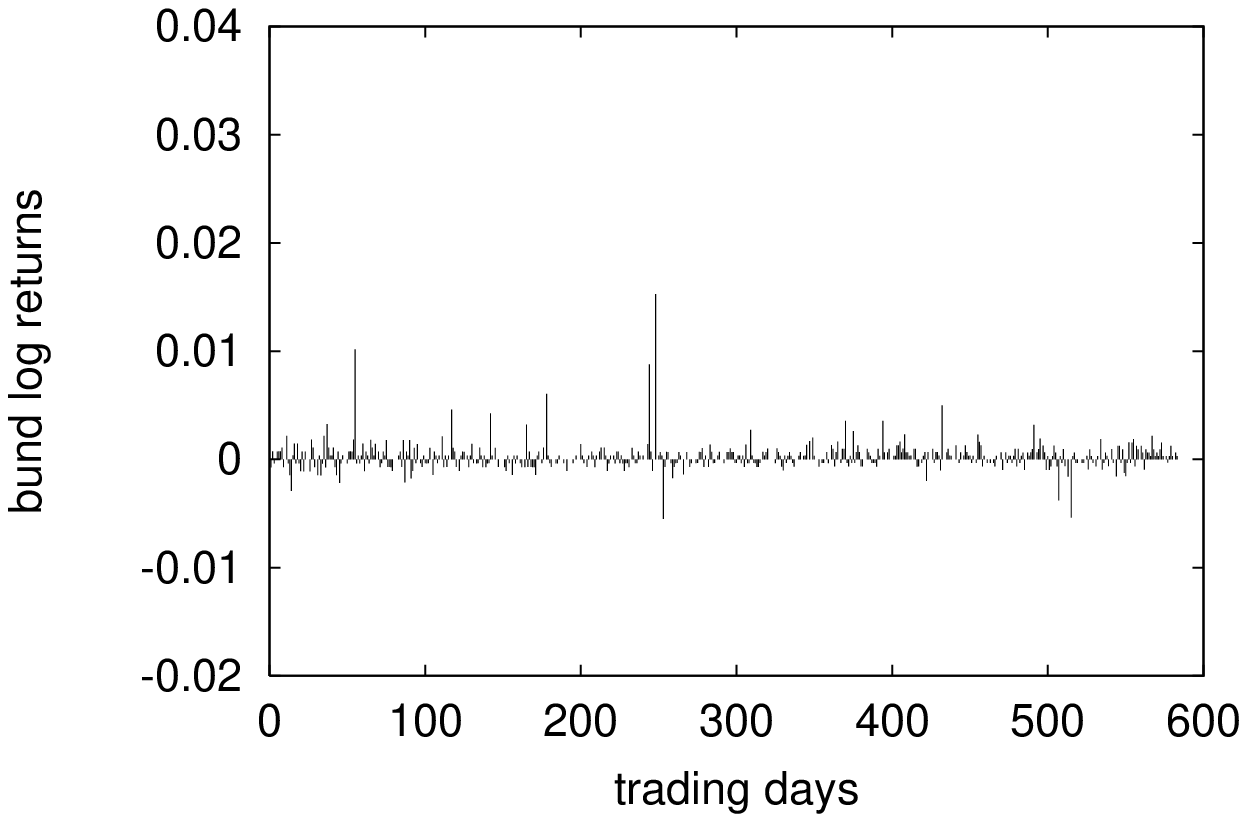, width=15pc}} \subfigure[]{\epsfig{file=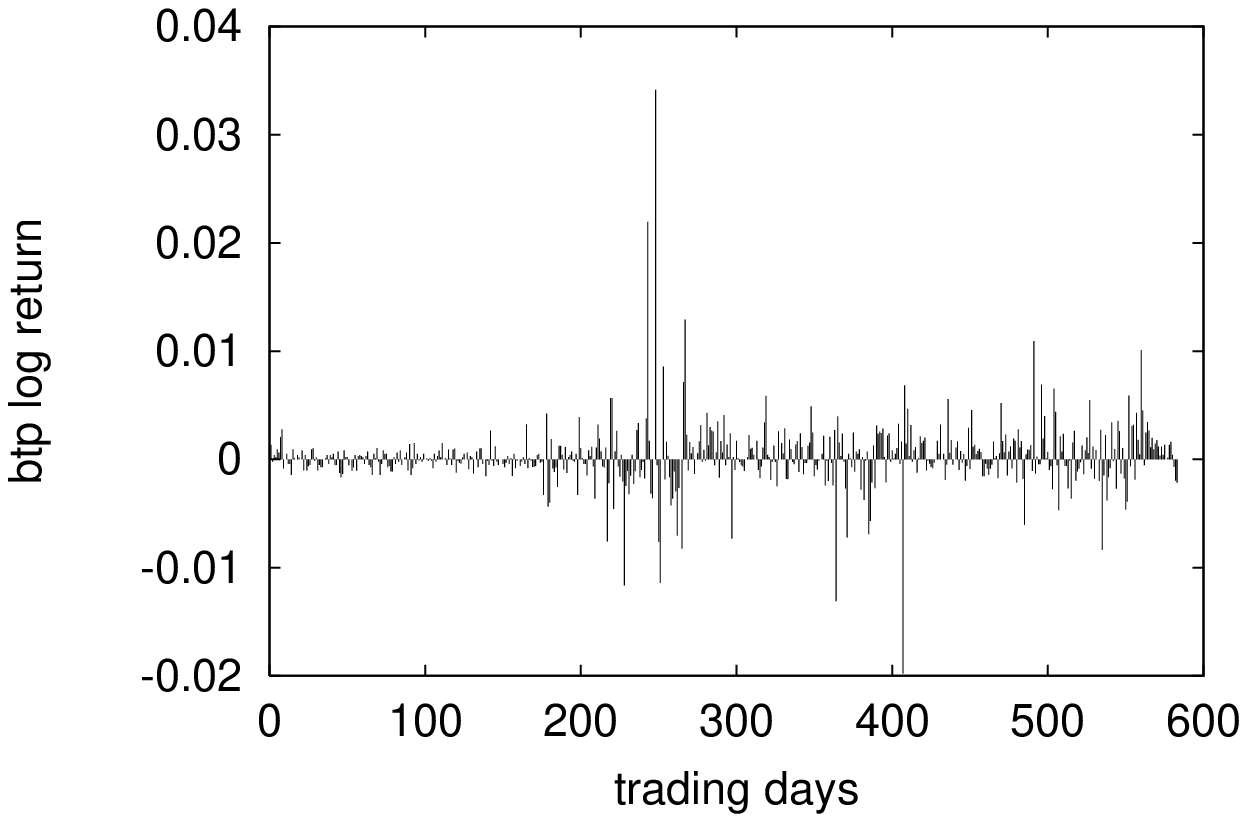, width=15pc}} \end{center} \SMALLCAP{Logarithmic returns for the \bund \ (a) and \btp \ (b) futures} \label{fig:returns} \end{figure}

\begin{figure} \begin{center} \subfigure[]{\epsfig{file=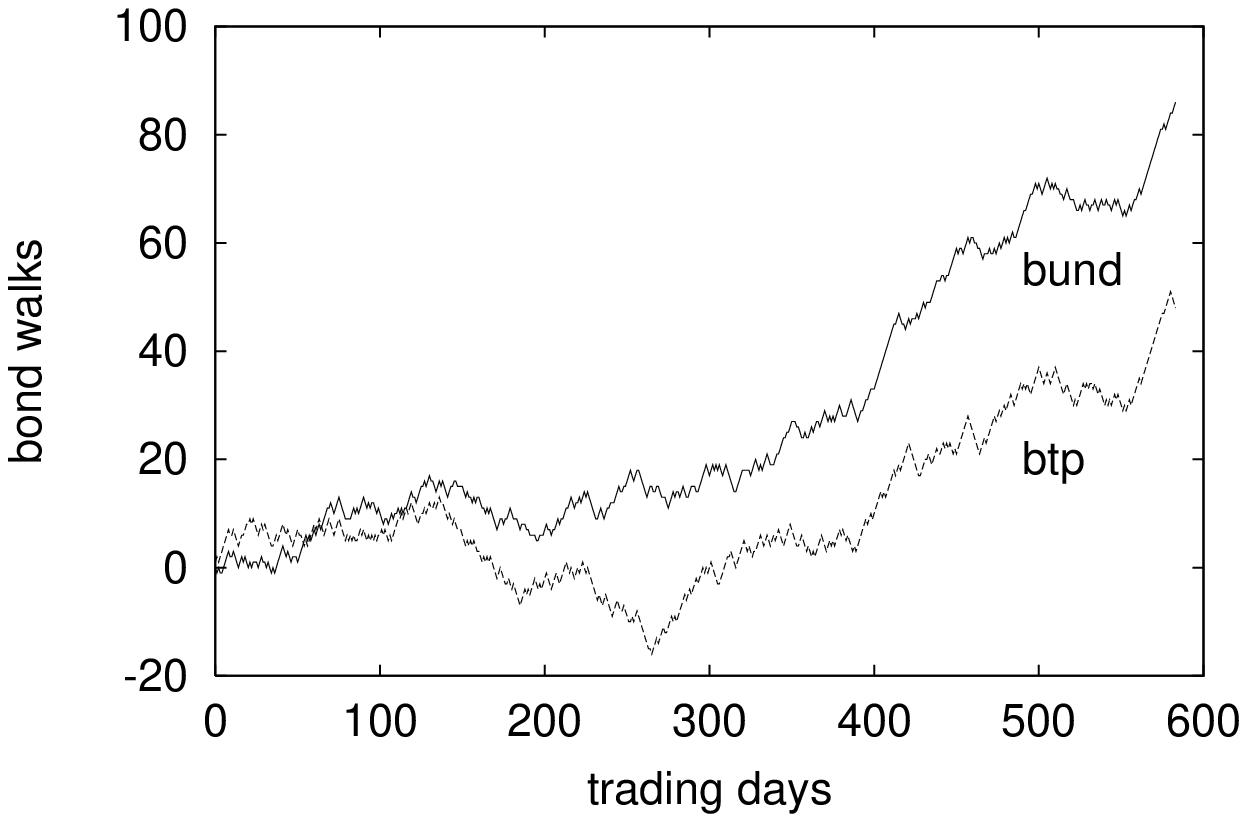, width=15pc}} \subfigure[]{\epsfig{file=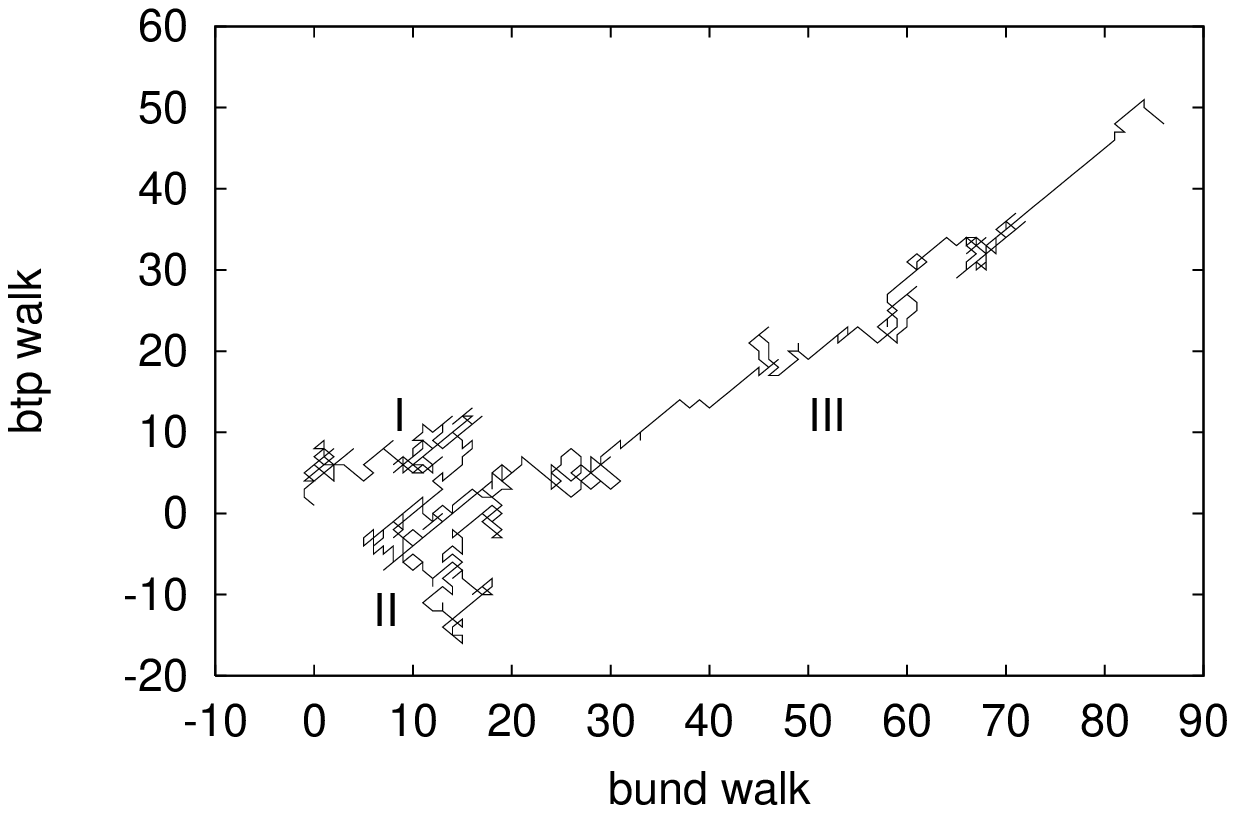, width=15pc}} \subfigure[]{\epsfig{file=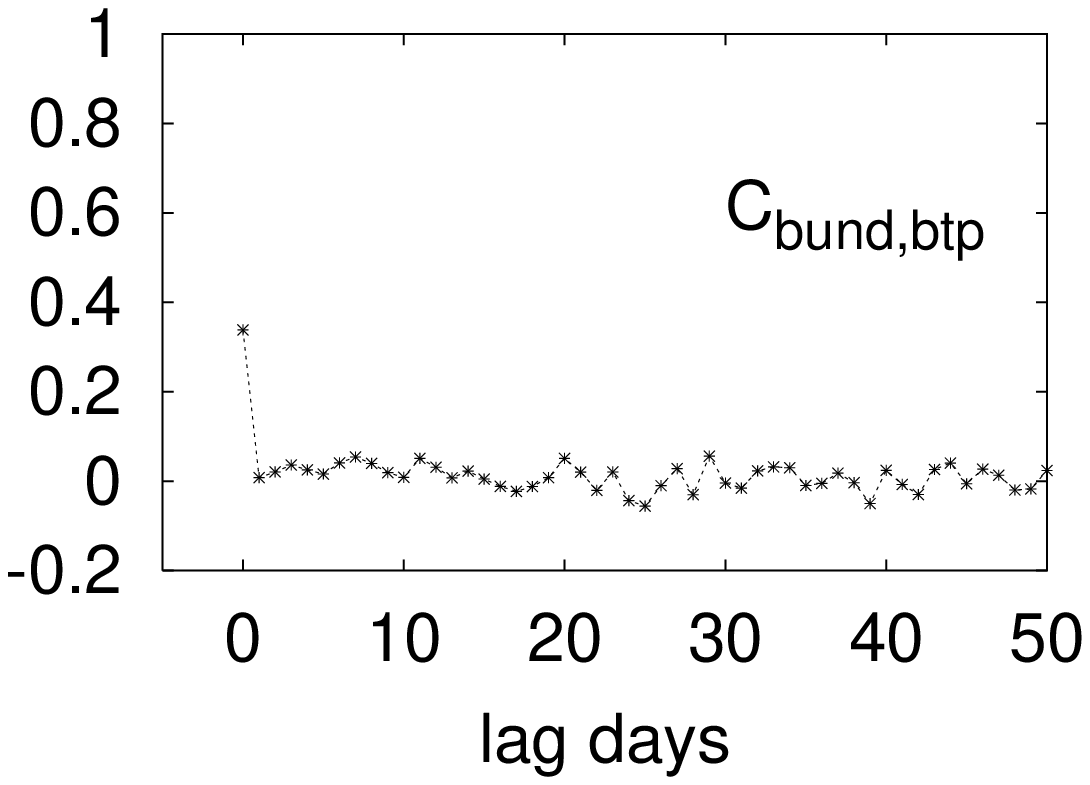, width=10pc}} \subfigure[]{\epsfig{file=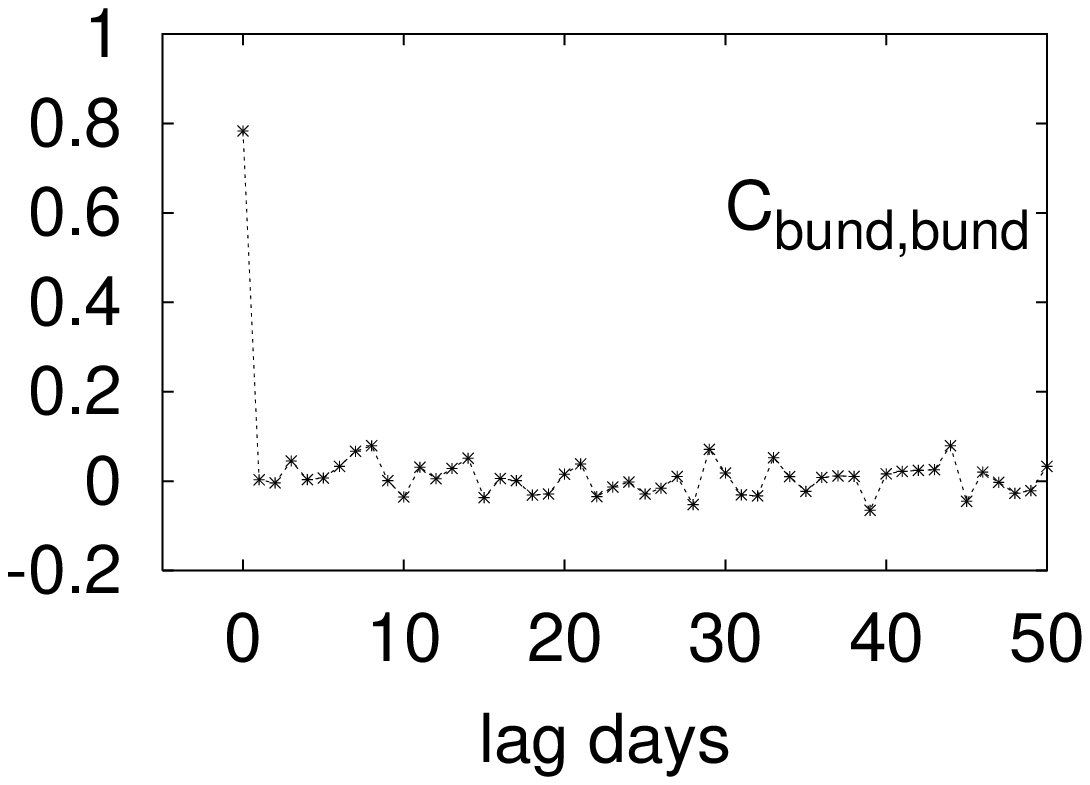, width=10pc}} \subfigure[]{\epsfig{file=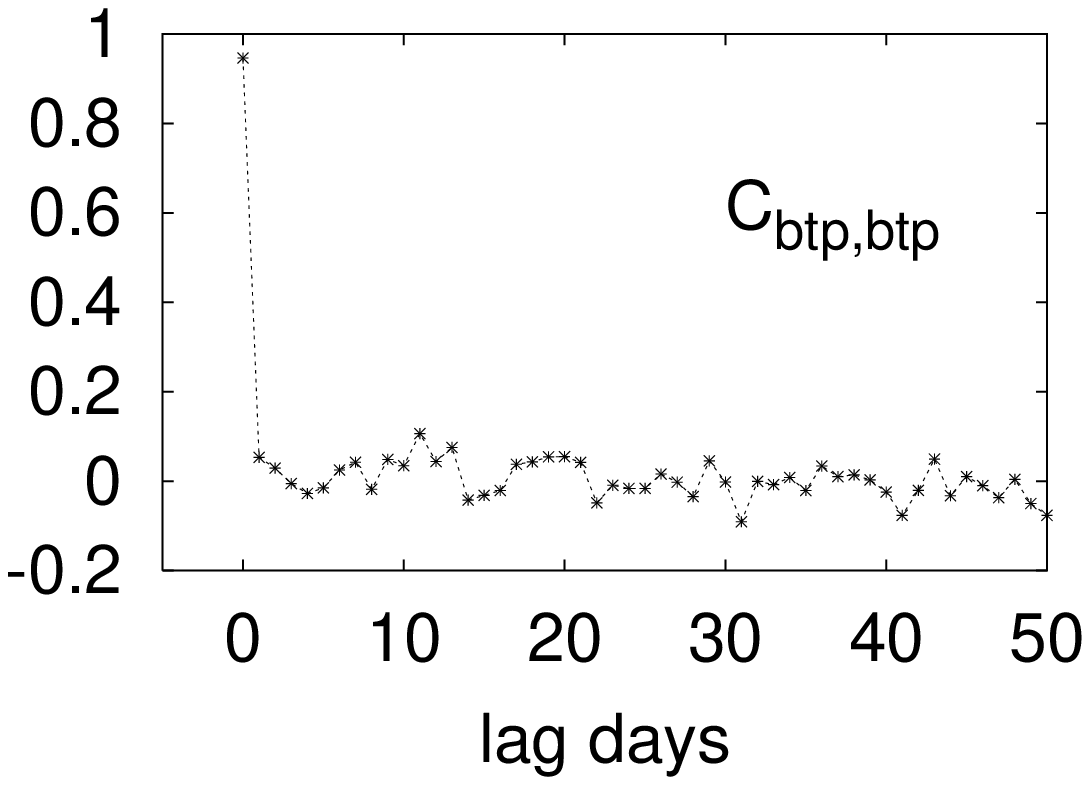, width=10pc}} \end{center} \SMALLCAP{(a) Displacement of the \bund \ walk (solid line) and the \btp \ walk (dashed line) futures; (b) \btp \ displacement vs \bund \ displacement random walk.
 Correlation functions for the bond walks: (c) \bund--\btp \ crosscorrelation; (d) \bund \ autocorrelation; (e) \btp \ autocorrelation} \label{fig:priceb} \end{figure}

\begin{figure} \begin{center} \subfigure[]{\epsfig{file=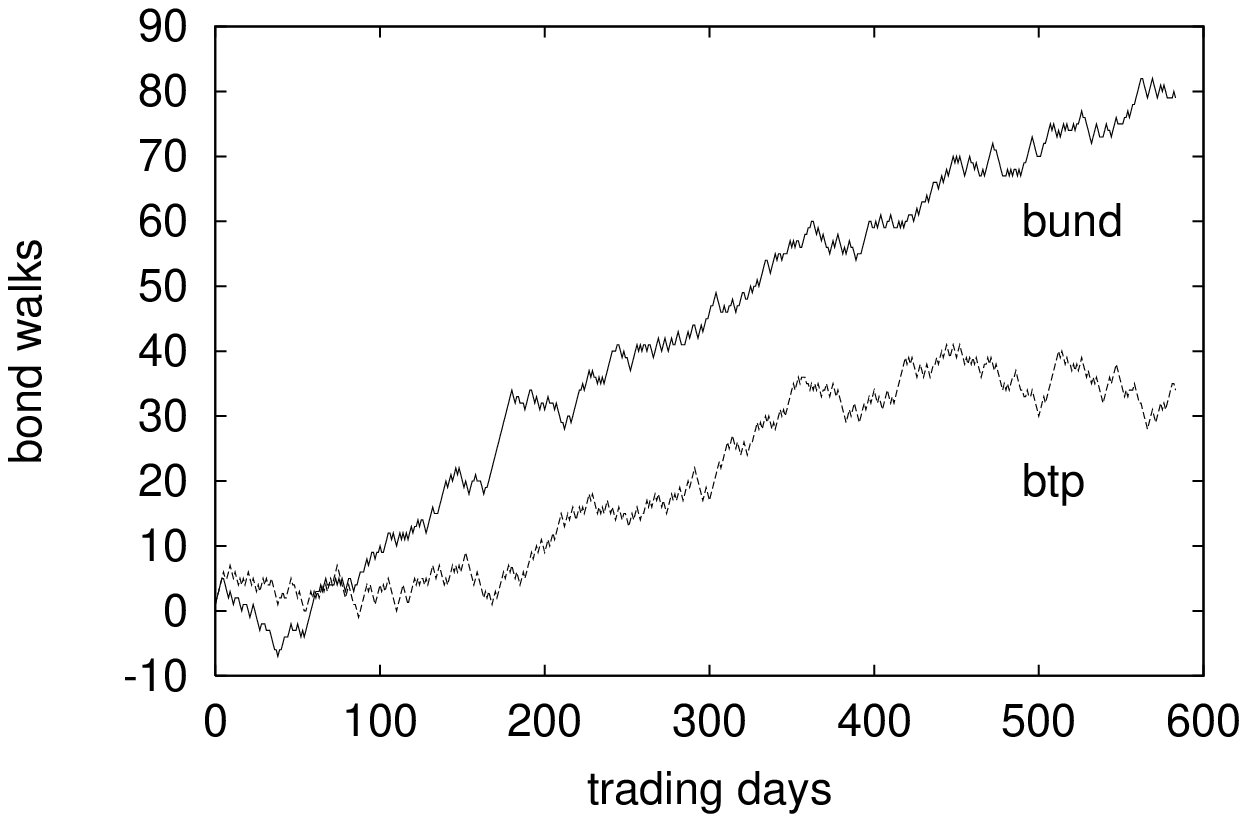, width=15pc}} \subfigure[]{\epsfig{file=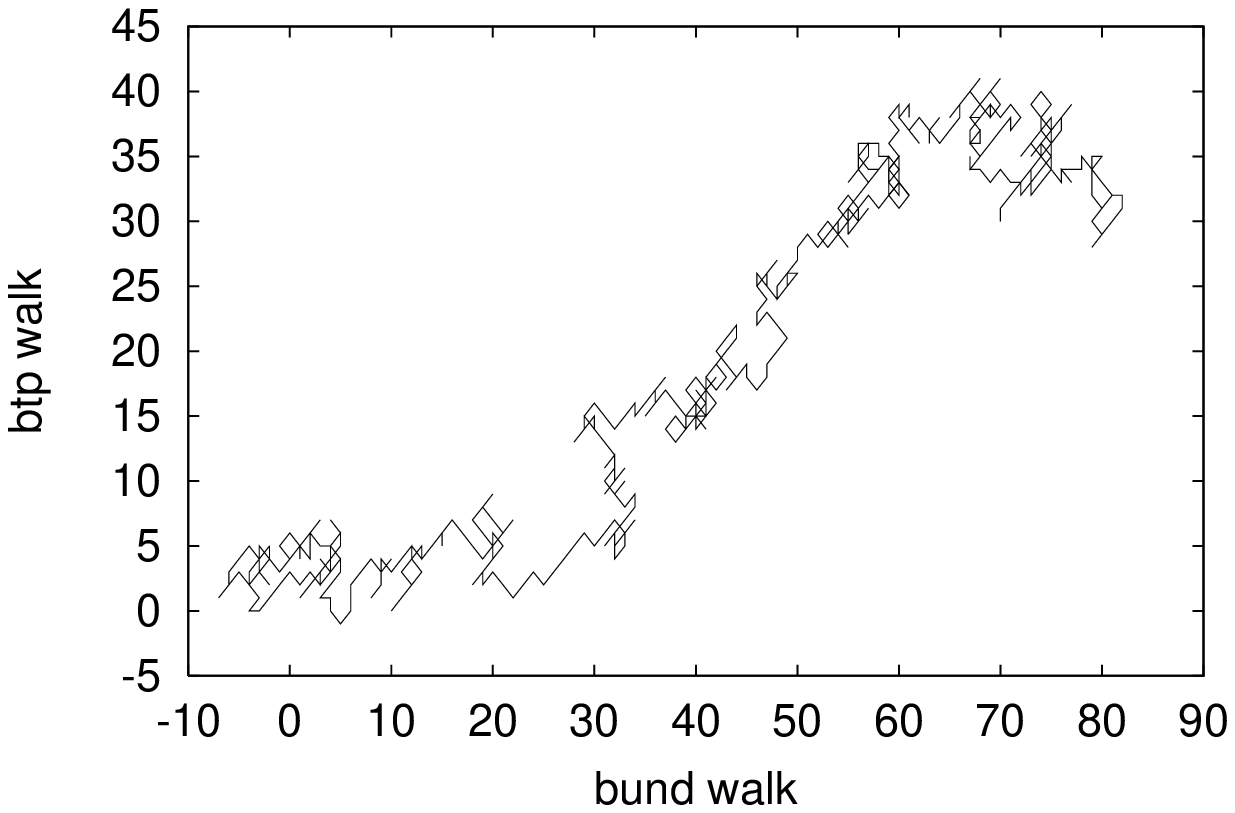, width=15pc}} \subfigure[]{\epsfig{file=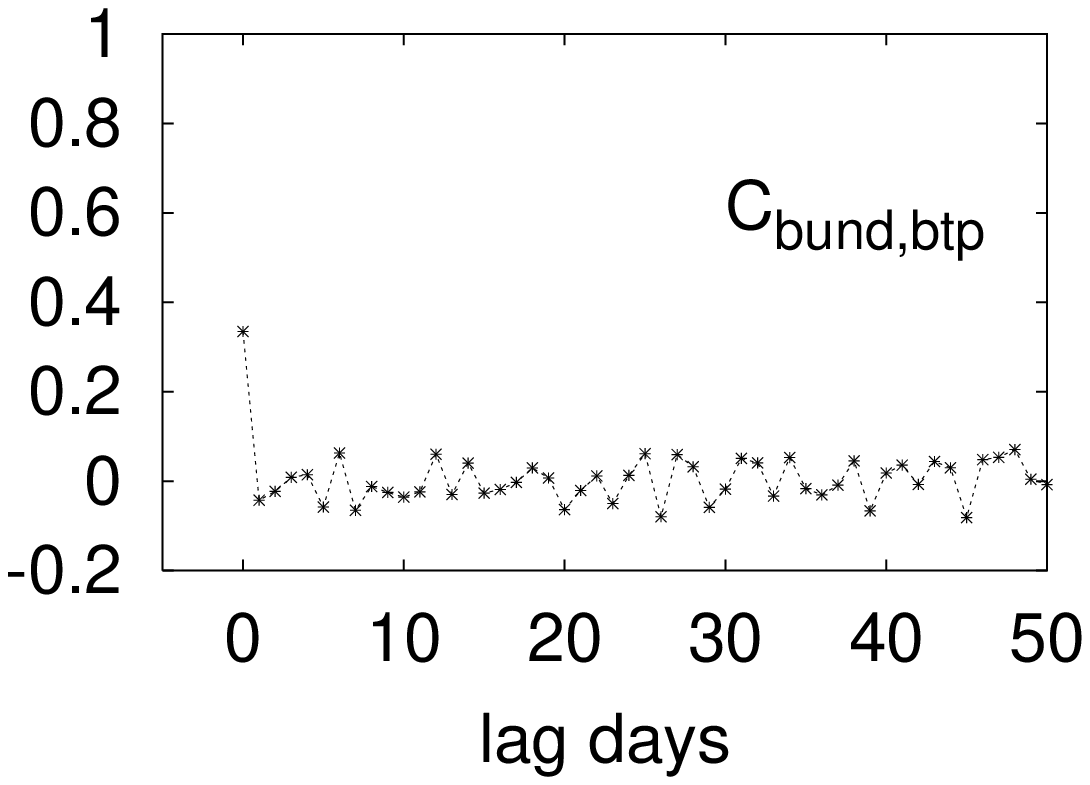, width=10pc}} \subfigure[]{\epsfig{file=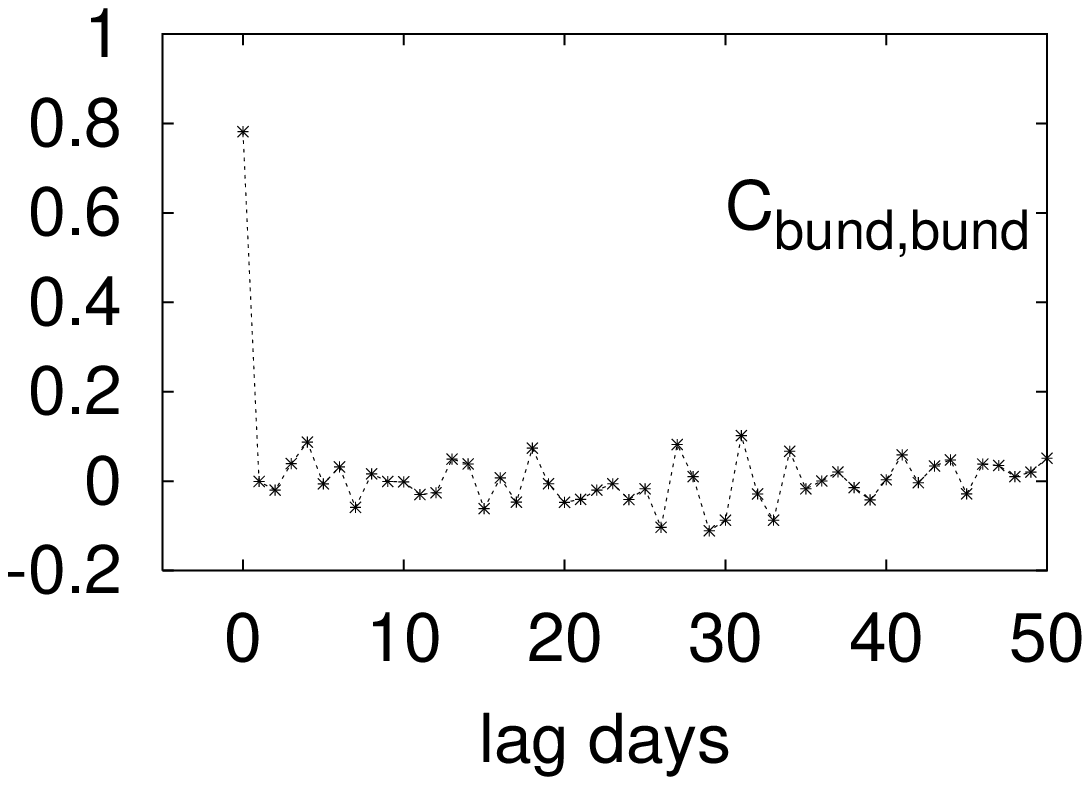, width=10pc}} \subfigure[]{\epsfig{file=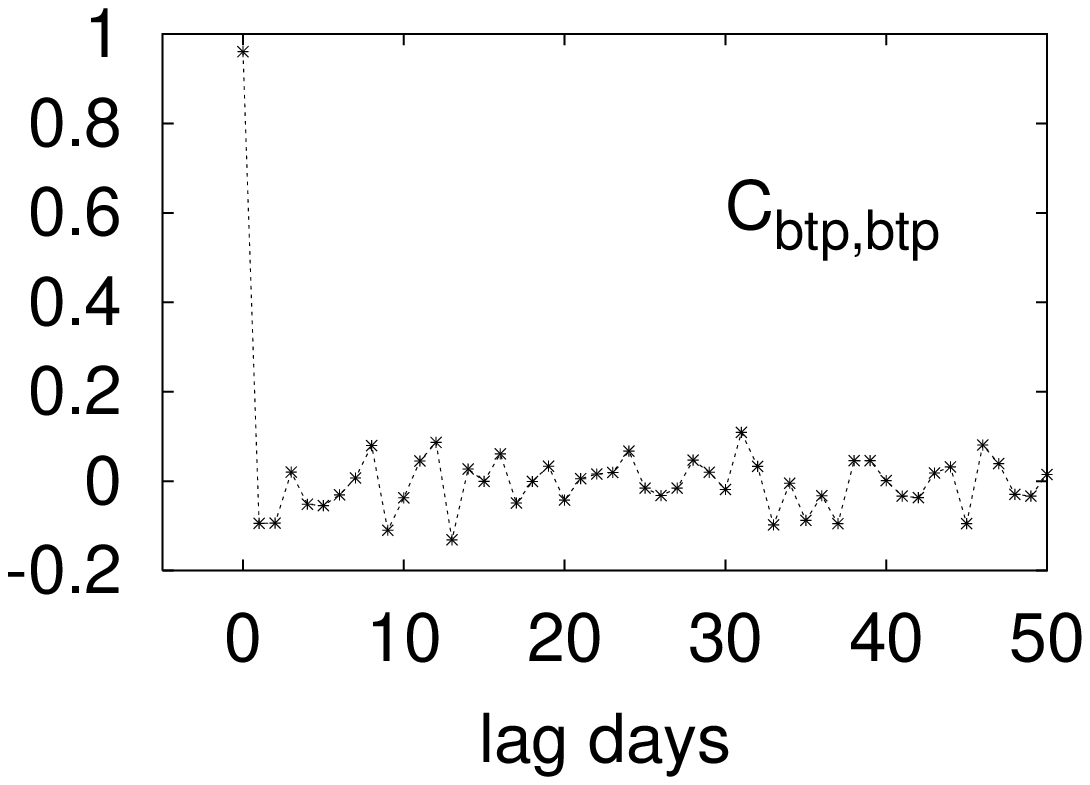, width=10pc}} \end{center} \SMALLCAP{Joint Monte Carlo displacements: (a) simulated \bund \ walk (solid line) and simulated \btp \ walk (dashed line); (b) simulated \btp \ displacement vs simulated \bund \ displacement random walk.
 Correlation functions for the joint simulated bond walks: (c)
 \bund--\btp \ crosscorrelation; (d) \bund \ autocorrelation; (e)
 \btp \ autocorrelation} \label{fig:mcpricebj} \end{figure}

In figure~\ref{fig:price}.a,  the time evolution of the \bund \ future as well as the \btp \ future closing prices is plotted as a function of the trading days, for the period October~1991---January~1994.
 At that time the \bundfuture \
 market opened earlier than the \btpfuture \ one.
 As a side remark, we notice that the volatility of the \btpfuture \ price is higher than that of the \bundfuture \
 price, which could be due to the lower liquidity of the \btp--contract market.
 In figure~\ref{fig:returns},  the logarithmic overnight returns
 \beann 
 r_b (n) = \log \lrb \frac{P_b^{\rm o} (n)}{P_b^{\rm c} (n-1)}
 \rrb, ~~~~~b=\mbox{\bund,\btp} 
 \eeann 
 are shown; $P_b^o$, and
 $P_b^c$ are the opening and the closing price.
 Here also, the greater volatility of the \btp \ contract is evident.
 \\ In this paper, we are not interested in the absolute value of the overnight variations, but only in their signs $u_b (n) = {\rm sign}_0 \lrb r_b (n) \rrb$, where ${\rm sign}_0$ coincides with the usual sign function except for the prescription ${\rm sign}_0 (0) \equiv 0$.
 Let us, now, introduce the {\em bond walk} displacement \cite{Scalas98,PBGHSSS92} as following:
 \beann 
 \ell_b (n) = \sum_{m=1}^n u_b (m) .
 \eeann 
 In figure~\ref{fig:priceb}.a,  the displacements $\ell_{\sbund}$ and $\ell_{\sbtp}$ are shown.
 This procedure visually enhances the correlation between the two price series, which becomes clearer in figure~\ref{fig:priceb}.b, where the two--dimensional random walk is now on a square lattice.
 The zero--lag value of the crosscorrelation between $u_{\sbund}$ and $u_{\sbtp}$ quantitatively measures how similar the two dynamics are.
 Indeed in figure~\ref{fig:priceb}.c, we find that the estimate of the crosscorrelation $C_{\sbund,\sbtp}(0)$ is significantly different from zero. Figures~\ref{fig:priceb}.d and \ref{fig:priceb}.e show that in each bond walk the autocorrelation
 function vanishes for any lag different from zero: Therefore there are neither long nor short range correlations in these walks.
 Correlations have been computed by using the unbiased estimator given in Ref. \cite{Marple87}.
 \begin{table} \begin{center} \begin{tabular}{l||c|c|c|c} &$u_{\sbtp} ~= -1$	&$u_{\sbtp} ~= 0$	&$u_{\sbtp} ~=+ 1$	&	\\ \hline \hline $u_{\sbund} ~= -1$		&	.22 (.68)	&	.01 (.03)	&	.09 (.29)	& .32	\\ \hline $u_{\sbund} ~= 0$		&	.09 (.47)	&	.02 (.09)	&	.09 (.44)	& .20	\\ \hline $u_{\sbund} ~= +1$		&	.12 (.25)	&	.02 (.04)	&	.34 (.71)	& .48	\\ \hline &	.43	&	.05	&	.52	& 1 \end{tabular} \end{center} \SMALLCAP{\label{table:prob}Contingency Table: Joint frequencies $f \lrb u_{\ssbund}~{\rm and}~u_{\ssbtp} \rrb$; in brackets conditional frequencies $f \lrb u_{\ssbtp}~{\rm given}~u_{\ssbund} \rrb $} \end{table} \\ In order to correctly describe the statistical correlations between the two bond walks, it is necessary to take into account the joint probability distribution or, equivalently, the disjoint probability distributions as well as the conditional probabilities.
 In Table~\ref{table:prob}, we give an estimate of the joint and conditional probabilities (in brackets) in terms of the empirical frequencies.
 In figure~\ref{fig:mcpricebj},  we show the results of a Monte Carlo simulation drawn from the joint probability distribution $p (u_{\sbund}~{\rm and}~u_{\sbtp})$.
 This simulation has been implemented as follows: At each tick, we randomly choose the \bund \ move according to the last column of Table~\ref{table:prob}; then, the \btp \ move is selected following Table~\ref{table:prob}. For instance, suppose that the extracted \bund \ move is upwards, then the probabilities for \btp \ move are given by the third row of Table~\ref{table:prob}.
 The results of the simulation are shown in figure~\ref{fig:mcpricebj}.
 In this case the zero--lag crosscorrelation value is significantly (and correctly) different from zero.
 \section{Gambling} \label{sec:gamb} 
\begin{figure} \begin{center} \subfigure[]{\epsfig{file=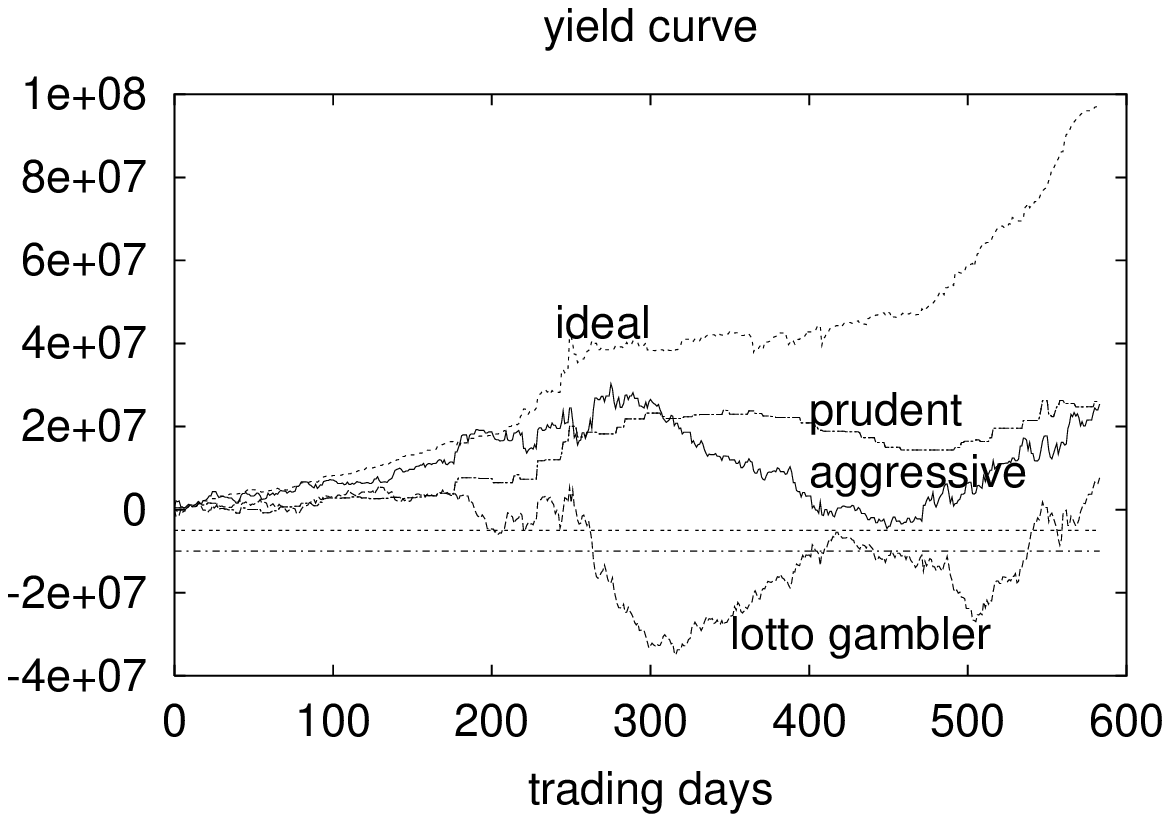, width=15pc}} \subfigure[]{\epsfig{file=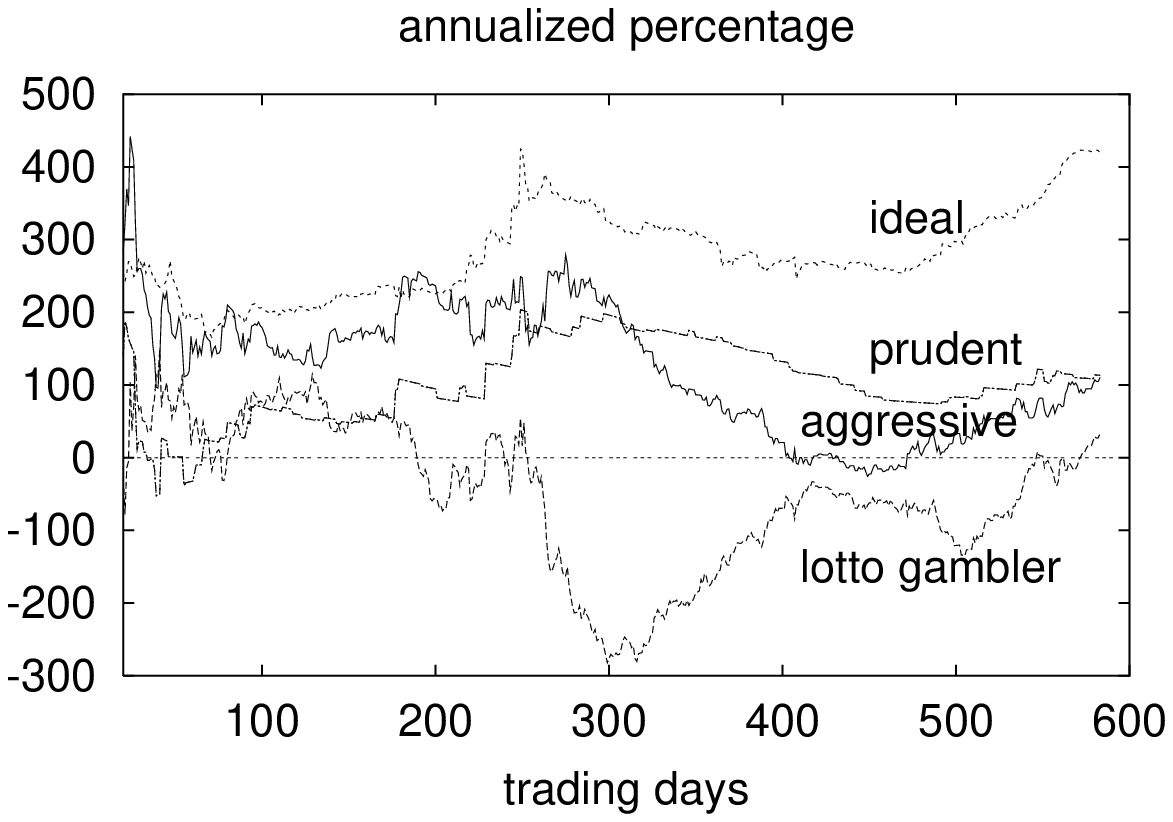, width=15pc}} \end{center} \SMALLCAP{(a) Yield Curves obtained using \bund \ overnight information; the horizontal lines are the initial and the maintenance margins for two contracts; (b) Annualized percentage
 obtained using \bund \ overnight information, the plot starts from the 20th trading day} \label{fig:yieldnewperc} \end{figure}

The previous analysis shows that the overnight signs of the two considered bond futures are crosscorrelated.
 One can now think to exploit this ``prior information'' to test the possibility of making profits.
 This is what we develop in this section, where the low (high) probability of opposite (equal) overnight signs (see Table~\ref{table:prob}) is used to build ``automatic investor'' profiles.
 Each profile corresponds to a precise investment strategy, fulfilling
 certain rules compatible with the future--market ones \cite{Hull96,Chance89,Duffie89}.
 At the first investment day, a {\em margin account} is created and filled with an {\em initial margin} for any contract opened \cite{Hull96}.
 In our case, on the first day, before the closing time, two \btpfuture \ contracts, a short and a long position\footnote{A short (long) position is a contract for selling (buying) a security at a certain future delivery date; in our case the security is a Treasury bond.}, are opened.
 Thus, at the beginning of each trading day, either the short or the long position is closed, depending on the chosen strategy.
 Before the closing time of the same day, the closed position is opened again\footnote{As a technical remark, we point out that, at the end of each day, all positions must be updated on the margin account for {\em marking to market}. The margin account must be fed when it becomes lower than the {\em maintenance margin}.}.
 \\ We call {\em aggressive} the profile for which, a positive (negative) \bundfuture \ overnight return implies the closure of the \btpfuture \ long (short) position; for zero returns no position is closed.
 The {\em prudent} automatic investor, on the other hand, closes the convenient morning position only if the \bundfuture \ return exceeds a certain threshold.
 If we define $u_b^\ve (n) = {\rm sign}_\ve \lrb r_b (n) \rrb$, where ${\rm sign}_\ve$ coincides with the usual sign function except for the prescription ${\rm sign}_\ve (x) \equiv 0$ for $|x| \le \ve$, we can use the $\ve$ parameter to characterize the ``aggressiveness'' of the investor.
 In figure~\ref{fig:yieldnewperc}.a,  the aggressive ($\ve = 0$) and a prudent ($\ve = 0.001$) investor performances are shown.
\\ It is not easy to place an order exactly at the opening price. However, suppose you know the \bundfuture \ sign variation half an hour before the opening time of the \btp \ market, then you can immediately phone your broker telling him/her what to do, thus increasing the possibility of closing your chosen position at the opening price.
 Indeed, in our calculation we
 assume that transactions are costless and happen exactly at the opening and closing prices. This assumption is quite strong when thinking to a real operation order.
 \\ In figure~\ref{fig:yieldnewperc}.a two other curves appear: The {\em lotto-gambler} and the {\em ideal} one.
 The lotto--gambler curve is built assuming the closure of the short position, the closure of the long one or neither of the two operations based on a trinomial probability distribution obtained by the past information on the \btpfuture \ contract.
 This algorithm is developed in the spirit of a `technical--analysis' attitude, where predictability of equity returns from past returns is assumed~\cite{BLLeB92}.
 \\ In formul\ae, \ the plotted yielding curves, $Y$, are defined as follows: \bea Y(n) = Y (n-1) - u(n) V_{\sbtp} \ds{\frac {P_{\sbtp}^{\rm c}(n) - P_{\sbtp}^{\rm o}(n)}{100} } , \label{eq:yielding} \eea where $u(\cdot)$ is $u_{\sbund}^{0.001} (\cdot)$ for~a~prudent~investor, $u_{\sbund} (\cdot)$ for~the~aggressive~investor, and ${\rm rnd} (\cdot)$ for~the~lotto~gambler~investor, and where ${\rm rnd}(n) = -1, 0, +1$ with probability $p_{-1} (n), p_0 (n), p_{1}(n)$ respectively. The probabilities $p_{(\cdot)}(n)$ are built using only the past information, \ie \ only using the distribution of $u_{\sbtp} (m < n)$.
 The quantity
 $V_{\sbtp}$ is the contract value fixed to $250,000,000$ {\sc itl} by {\sc liffe}.			 \\ In the ideal profile, we exploit the out--of--the--rule possibility of opening a \btpfuture \ position --at the closing time of the previous day-- in the time between the opening of the \bund \ market and the opening of the \btp \ market, and of closing the same position immediately after this time; the position will be long (short), if the \bund \ overnight is positive (negative) and no operation is done for zero overnight returns.
 \\ In figure~\ref{fig:yieldnewperc}.b, the plot of the annualized percentages is presented 
 \beann
 \Pi(n) = \frac\alpha n \lrb \frac {Y(n)} {Y(0)} - 1 \rrb , \label{eq:percentage} 
 \eeann 
 where $\alpha$ is given by the product of 254 (trading days per year) and 100 (percentage magnification) and $Y(0)$ equals to the initial margin.
 To open a future position, only this initial margin is necessary.
 \\ Though not practically achievable, the ideal profile is the realization which better takes into account the presence of correlations, giving yields, on the long run, four time greater than the other profiles. The explanation of this fact is as follows: The ideal profile is the only one where the information contained in the overnight crosscorrelation is fully exploited.
 In the other cases, this information is only partially used due to market rules.
 \section{Discussion and Conclusions} \label{sec:concl} In this paper, we have studied the correlations between bond walks for \bund \ and \btp \
 time series.
 We have found a situation similar to the one in experiments with correlated photons \cite{EPR35}.
 If the two walks are separately analyzed, their statistical properties can be described by random walks with trinomial probability transitions.
 However, if we consider crosscorrelations, we find that the two walks are not independent one from the other.
 In this case, of course, there are neither quantum entanglements nor non--local quantum effects. It is likely that the operators in the \btpfuture \ market simply check the \bundfuture \ overnight sign and behave accordingly.
 \\ In the second part of the paper, we have investigated the possibility of exploiting the above correlation in order to realize a profit.
 Various strategies have been explored and it seems that, using the information contained in overnight correlations could lead to
 non--irrelevant yields.
 Indeed, nowadays the two markets open at the same time, thus eliminating these profit potentialities.
\\ Which is the origin of the behaviour of the yield curves? In equation~(\ref{eq:yielding}), there is a profit if there are positive correlations between the two bond walks. In the case of the aggressive investor, a negative correlation always determines a loss, whereas this is not the case in the prudent case. Therefore, in periods of strong positive correlations both strategies lead to profits, which are greater in the aggressive case; in periods characterized by weaker correlations, there can be either profits or losses depending on the absolute value of price variations. Finally, in a period of anticorrelations, the aggressive investor systematically loses money, whereas the prudent investor loses money only if \bund \ price variations exceed a threshold.
\\ One may ask whether the observed positive correlations giving rise to profits are due to random fluctuations.
 If one takes into account the full data set (N = 584 points), a two--factor linear regression analysis of the data plotted in figure~3.b gives a correlation coefficient $r = 0.89$. The null hypothesis of no correlations can be checked by a $t$-Student's statistics test \cite{kendall1967} and it is rejected
even for a 99.5 \% confidence interval, being $t=49$. However, a careful inspection of figure~3.b shows that
three definite regions can be distinguished:
Region I, including the first 150 points (from 19/09/1991 to 23/04/1992),
region III covering the last 264 points (from 22/12/1992 to 11/01/1994), and
region II in the middle. In region I and III, positive correlations are
strong. 
In particular, in region III the correlation coefficient is $r=0.98$
with $t=79$, whereas in region I $r=0.63$ and $t=10$. In both regions the
null hypothesis of no correlation is rejected for a 99.5 \% confidence interval.
In region II, on the contrary, the null hypothesis cannot be rejected at a
99.5 \% confidence level. In fact, $r=0.18$ and $t=2.5$.
\\ An intriguing point is the origin of the observed correlations;
it is also interesting to understand why there is a temporal window 
of weaker correlations, during 1992.
One reason for the presence of positive correlations is the strong link
between the German and the Italian bond--markets. Indeed the Italian an German
economies were deeply interwoven, and the values of the two currencies were related by the European Exchange--Rate Mechanism (ERM). 
As for the second question, one should notice that, due to speculative
pressure, the Italian currency had to be devaluated thus leaving the ERM in 1992.
\\ The method described in this paper can be easily generalized to investigate multiple correlations between assets. 
For instance, correlations of \tbond \ (U.S. government bonds) futures, \bund \ and \btp \
futures could be considered. 
Moreover, it is possible to use zero--lag two--point crosscorrelations of asset walks to measure distances in a
hierarchical analysis of markets~\cite{Mantegna97b}.

\section{Aknowledgments}
\label{sec:aknow}

We gratefully acknowledge fruitful discussion with Marina Resta. We are indebted to Massimo
Riani for discussion, support, and encouragement.
The \bund-- \ and \btpfuture \ data are available at {\sc liffe} ({\tt www.liffe.com}).


\begin{thebibliography}{10}
{\small

\bibitem{AAP88}
J.P. Bouchaud and M.~Potters,

\newblock Th\'eorie des Risques Financiers (Al\'ea Saclay, Paris, 1997);
J.~Kertesz and I.~Kondor, eds.,
\newblock Econophysics: an Emerging Science (Kluwer Academic Publishers,
  Dordrecht, Netherlands, 1998);
R.N. Mantegna and H.E. Stanley,

\newblock Scaling Concepts in Finance (Cambridge University Press, 1998).

\bibitem{Dumbar98}
N.~Dumbar,
\newblock New Scientist 158 (1998);
J.M. Pimbley,

\newblock Physics Today 50 (1997) 42--46;
A.L. Robinson,
\newblock Physics Today 47 (1994) 55--56;
G.~Stix,
\newblock Scientific American 278 (1998) 70--75.

\bibitem{Mantegna97b}
R.N. Mantegna,
\newblock Degree of correlation inside financial market,
\newblock in: Applied Nonlinear Dynamics and Stochastic Systems near the
  Millenium, J.B. Kadtke and A.~Bulsara, eds. (AIP press, 1997);
R.N. Mantegna,
\newblock preprint {\tt cond-mat/9802256} (1998).

\bibitem{Scalas98}
E.~Scalas,
\newblock Physica A 253 (1998) 394--402.

\bibitem{PBGHSSS92}
C.K. Peng, S.V. Buldyrev, A.L. Goldberger, S.~Havlin, F.~Sciortino, M.~Simons
  and H.E. Stanley,
\newblock Nature 356 (1992) 168--170.

\bibitem{Marple87}
S.L. Marple,
\newblock Digital Spectral Analysis with Applications (Prentice--Hall, Inc.,
  1987).

\bibitem{Hull96}
J.C. Hull,
\newblock Options, Futures, and Other Derivatives,
\newblock 3rd Ed. (Prentice--Hall, Inc., Englewood Cliffs, NJ, 1996).

\bibitem{Chance89}
D.~Chance,
\newblock An Introduction to Options and Futures (Dryden Press, Orlando, FL,
  1989).

\bibitem{Duffie89}
D.~Duffie,
\newblock Futures Markets (Prentice--Hall, Inc., Englewood Cliffs, NJ, 1989).

\bibitem{BLLeB92}
W.A. Brock, J.~Lakonishok and B.D. LeBaron,
\newblock The Journal of Finance XLVII (1992) 1731--1764.

\bibitem{EPR35}
A.~Einstein, B.~Podolsky and N.~Rosen,
\newblock Physical Review Letters 47 (1935) 777--780;
J.S. Bell,
\newblock Physics 1 (1964) 195--200;
A.~Aspect, J.~Dalibard and G.~Roger,
\newblock Physical Review Letters 49 (1982) 1804--1808.
}

\bibitem{kendall1967}
M.G. Kendall and A. Stuart, 
\newblock The Advanced Theory of Statistics, Vol. II, (Charles Griffin \& Company Ltd, London, UK, 1967) 

\end{thebibliography}
\end{document}